# A Method for Characterizing Communities in Dynamic Attributed Complex Networks

GünceKezibanOrman[#*], Vincent Labatut[*], Marc Plantevit[+], Jean-François Boulicaut[#]

[#] *INSA-Lyon, CNRS, LIRIS,Université de Lyon*
*UMR5205, F-69621, Lyon, France*
[4]`jean-francois.boulicaut@insa-lyon.fr`

[*]*Computer Engineering Department, Galatasaray University*
*Ortaköy 34349, Istanbul, Turkey*
[1]`korman@gsu.edu.tr`, [2]`vlabatut@gsu.edu.tr`

[+]*Université Lyon 1,CNRS, LIRIS, Université de Lyon*
*UMR5205, F-69622, Lyon, France*
[3]`marc.plantevit@liris.cnrs.fr`

*Abstract*—**Many methods have beenproposed to detect communities, not only in plain, but also in attributed, directed or even dynamic complex networks. In its simplest form, a community structure takes the form of apartition of the node set. From the modeling point of view, to be of some utility, this partition must then be characterized relatively to the properties of the studied system. However, if most of the existing works focus on defining methods for the detection of communities, only very few try to tackle this interpretation problem. Moreover, the existing approaches are limited either in the type of data they handle, or by the nature of the results they output. In this work, we propose a method to efficiently support such a characterization task. We first define a sequence-based representation of networks, combining temporal information, topological measures, and nodal attributes. We then describe how to identify the most emerging sequential patterns of this dataset, and use them to characterize the communities. We also show how to detect unusual behavior in a community, and highlight outliers. Finally, as an illustration, we apply our method to a network of scientific collaborations.**

## I. INTRODUCTION

Complex networks have become very popular as amodeling tool during the last decade, because they help to better understand the intrinsic lawsand dynamics of complex systems. A typical plain network contains only nodes and links between them, but it is possible to enrich it with different types of data: link orientation and/or weight, temporal dimension, attributes describingthe nodes or links, etc. This flexibility allowed to use complex networks to study real-world systems in many fields: sociology, physics, genetics, computer, etc.[1].

The complex nature of the modeled systems leads to the presence of non-trivial topological properties in the corresponding networks. Among them, the *community structure* is one of the most common and most studied. Intuitively, we can define a *community* as a group of nodes which is densely interconnected relatively to the rest of the network[2].However, in the literature, this notion is formalized in many different ways[3]. There are actually hundreds of algorithms for detecting community structures, characterized by the use of different formal definitions of what a community is, and/or relying on different processes. Among the principles these approaches are based upon, one can cite: modularityoptimization, similarity matrix clustering, data compression, statistical significance, information diffusion, clique percolation, etc. (c.f. [3]for a detailed review). Most of the existing methods deal with plain complex networks, but new methods were gradually introduced to handle richer networks: first link directions and weights, then time, and more recently node attributes[4-8].

Although the algorithms differ in terms of nature of the detected communities, algorithmic complexity, result quality and other aspects[3], their output can always be basically described as a list of node groups. More specifically, in the case of mutually exclusive communities, it is a partition of the set of nodes. From an applicative point of view, the question is then to make sense of these groups relatively to the studied system. In other terms, for the community structure to be useful, it is necessary to interpret the detected communities.This problem is extremely important from the end user's perspective. And yet, almost all works in the field of community detection concern the definition of detection tools, and their evaluation in terms of performance. Only a very few works try to tackle the problem of characterizing and interpreting the communities.

Authors historically interpreted their data manually[9, 10]but this somewhat subjective approach does not scale well on large networks. More recently, several authors used topological measures to characterize community structures. In [11],Lancichinetti*et al.* visually examine the distribution of some community-based topological measures,both at local and intermediary levels. Their goal is to understand the general shape of communities belonging to networks modeling various types of real-world systems. In[12], Leskovec*et. al*propose to study the community structure as a whole, by considering it at various scales, thanks to a global measure called conductance. These two studies are interesting, because they try to describe the community structure of plain networks. However, it should be noted the resulting observations are quite general, in the sense communities are studied and characterized collectively, to identify trends in a network[12], or even a collection of networks[11].

In order to characterize each community individually, some authors take advantage of the information conveyed by nodal attribute, when it is available. In [13], the authors propose a statistical method to characterize the communities in terms of the over-expressed attributes found in the elements of the community. In[14], authors interpret the communities of a social attributed network. They use statistical regression and discriminant correspondence analysis to identify the most characteristic attributes of each community.Both studies are valuable, however they do not take advantage of the available topological measures to enhance the interpretation process.

As mentioned before, there are methods taking advantage of both relational (structure) and individual (attributes) information to detect communities. It seems natural to suppose the results they output can be used for interpretation purposes. For example, in [5], the authors interpret the communities in terms of the attributes used during the detection process; and in [15], the authors identify the top attributes for each identified community. However, the problem with these community detection-based method is that the notion of community is often defined procedurally, i.e. simply as the output of the detection method, without any further formalization. It is consequently not clear how structure and attributes affect the detection, and hence the interpretation process. All these methods additionally rely on the implicit assumption of community homophily. In other words, communities are supposed to be groups of nodes both densely interconnected and similar in terms of attributes. To our knowledge, no study has ever shown this feature was present in all systems, or even in all the communities of a given network. It is therefore doubtful those methods are general enough to be applied to any type of network.

In this work, we seethe interpretation problem as independent from the approach used for community detection, and we try to propose a method allowing to tackle the limitations of the existing approaches. Based on the observation that attributes allow to improve the interpretation of communities, we propose to enhance it further by considering temporal information, in addition to structure and attributes. Moreover, to obtain intelligible results, we want to explicitly identify which parts of the structural information are relevant to the interpretation process. For this matter, we detect common changes in topological features and attribute values over time periods, in dynamic attributed networks. More precisely, we aim at finding the most representative emerging sequential patterns for each community. These patterns can then be used for both characterizing the community, and identify outliers, i.e. node with non-standard behavior. The emergingpatterns represent the general trend of nodes in the considered community, whereas the outliers can correspond to nodes with a specific role in the community, or node located at itsfringe. We illustrate our proposal on a dynamic co-authorship network extracted from DBLP.

Our first contribution is to consider community characterization as a specific problem, distinct from community detection. In particular, it should be independent from the method used to detect communities, rely on an easily replicable systematic approach, and be as automated as possible. Our second contribution is the introduction of a new representation of dynamic attributed networks. It takes the form of a database containing sequences of node topological measures, attributes and community information, for several time slices. Such sequences were used for the representation of natural data before[16], but not for the networks. Our third contribution is the definition of a method taking advantage of this representation to extract sequential patterns able to characterize the communities. Finally, our last contribution is to illustrate our method by applying it to a real-world network.

The rest of this article is organized as follows. In section II, we give the preliminary definitions needed to describe our method. In section III, we specify the problem and explain in detail our interpretation method. In section IV, we present our experimental results obtained on the DBLP data. Section V discuss our work and presents its possible extensions.

## II. PRELIMINARY DEFINITIONS

In this section, we first define different network-related concepts needed to introduce our method. We also describe the topological measures we use in our experiment.

### A. Network and Community Structure

We define a *dynamicnetwork* $G = \langle G_1, \ldots, G_\theta \rangle$ as a sequence of chronologically ordered time slices. Each *time slice* corresponds to a separated subnetwork $G_t$ ($1 \leq t \leq \theta$), which represents the connections between the nodes for a given time interval. Moreover, the networks we consider are*attributed*, meaning their nodes are described by some individual attributes. We therefore note a time slice $G_t = (V, E_t, A)$, where $V$ is the set of nodes, $E_t \subseteq V \times V$ is the set of links and $A$ is the set of node attributes. In what we call a*dynamicattributed network*, the node and attribute sets are the same at each time slice, whereas the link repartition and attribute values can change. We note $|V| = n$ the number of nodes.We only deal with undirected graphs,so we consider all pairs $(w, v)$ to be unordered. We also define the following adjacency function $f_t$:

$$f_t(w, v) = \begin{cases} 1 \text{ if } (w, v) \in E_t \\ 0 \text{ if } (w, v) \notin E_t \end{cases} \quad (1)$$

where$f_t(w, v)$directly depends on the presence (value 1) or absence (value 0) of a link between nodes $w$ and $v$ at time $t$.

We define the *global weighted network* $\mathcal{G} = (V, \mathcal{E})$ associated to a dynamic network $G$as its integration over time. The link set of $\mathcal{G}$ contains all links appearing through time $\mathcal{E} = \cup_{t=1}^{\theta} E_t$, whereas the node set is the same than for $G$, since it is fixed. We define a weight function $f$ summing $f_t$ over time: the weight of a link $(w, v)$from$\mathcal{E}$ is defined as:

$$f(w, v) = \sum_{1 \leq t \leq \theta} f_t(w, v) \qquad (2)$$

A *community structure* of a network is a partition of its node set, and each part of this partition is called a community. Here, we work with a static community structure $\mathcal{C}$ of $\mathcal{G}$, so a partition of $V$, and we note the communities $C_c$ ($1 \leq c \leq \lambda$), for $N$ distinct communities. Moreover, we define $C(v)$, the function associating a node $v$ to its community in $\mathcal{G}$. The *community size* of a given community $C_c$ is $|C_c|$, i.e. the number of nodes it contains.

### B. Topological Measures

A *topological measure* quantifies the structural properties of the network or its components. Here, we focus on five nodal measures: internal degree, local transitivity, within module degree, participation coefficient and embeddedness. We process each of these measures for each node, and at each time slice. The degree and local transitivity are both local measures, whereas the others are community-related.

We first note $N_t(v) = \{w \in V : (v, w) \in E_t\}$ the *neighborhood* of node $v$ at time $t$, i.e. the set of nodes connected to $v$ in $G_t$. The *degree* $d_t(v) = |N_t(v)|$ of a node is the cardinality of its neighborhood, i.e. its number of neighbors, at time slice $t$.

We define the *internal neighborhood* of a node $v$ at time $t$ as the subset of its neighborhood located in its community: $N_t^{int}(v) = N_t(v) \cap C(v)$. The *internal degree* $d_t^{int}(v) = |N_t^{int}(v)|$ is defined similarly to the degree, as the cardinality of the internal neighborhood, i.e. the number of neighbors the node $v$ has in its community at time $t$.

The *local transitivity* corresponds to the ratio of existing to possible triangles containing $v$ in $G_t$:

$$T_t(v) = \frac{|\{(q, w) \in E_t : q \in N_t(v) \wedge w \in N_t(v)\}|}{d_t(v)(d_t(v) - 1)/2} \qquad (3)$$

In this ratio, the numerator corresponds to the observed number of links between the neighbors of $v$, whereas the denominator is the maximum possible number of such links.

The *within module degree* and *participation coefficient* are two measures proposed by Amaral & Guimerà[17] to characterize the community role of nodes. The *within module degree* is defined as the $z$-score of the internal degree:

$$z_t(v) = \frac{d_t^{int}(v) - \mu\left(d_t^{int}, C(v)\right)}{\sigma\left(d_t^{int}, C(v)\right)} \qquad (4)$$

Here, $\mu$ and $\sigma$ denote the mean and standard deviation of $d_t^{int}$ over all nodes belonging to $C(v)$, respectively. It expresses how much a node is well connected to other nodes in its community, relatively to this community. In[17], the authors distinguish nodes depending on whether their $z_t$ is above or below a limit of 2.5. If $z_t \geq 2.5$, the node is considered to be a community hub, because it is significantly more connected to its community than the other members of the same community. If $z < 2.5$, the node is said to be a community non-hub.

The *participation coefficient*, introduced in the same study[17], is based on the notion of *community degree* $d_t^c(v) = |N_t(v) \cap C_c|$, which represents the number of links a node $v$ has with nodes belonging to community $C_c$. Incidentally, one can see the internal degree is a specific case of community degree, for which $C_c = C(v)$. This measure is formally defined as:

$$P_t(v) = 1 - \sum_{1 \leq c \leq \lambda} \left(\frac{d_t^c(v)}{d_t(v)}\right)^2 \qquad (5)$$

where $\lambda$ is the number of communities in $\mathcal{G}$. $P$ characterizes the distribution of the neighbors of a node over all communities. More precisely, it measures the heterogeneity of this distribution. It gets close to 1 if all the neighbors are uniformly distributed among all the communities and 0 if they are all gathered in the same community.

The *embeddedness* represents the proportion of neighbors of a node belonging to its own community[11]. Unlike the within module degree, the embeddedness is normalized with respect to the node, and not the community.

$$e_t(v) = \frac{d_t^{int}(v)}{d_t(v)} \qquad (6)$$

A *node descriptor* is either any of these five topological measures explained above, or a node attribute from $A$. Let $D = \{D_1, D_2, \dots, D_k\}$ be the set of all descriptors. Each descriptor from $D$ can take one of several discrete values, defined in its *domain* $\mathfrak{D}_i$ ($1 \leq i \leq k$).

### III. CHARACTERIZATION METHOD

We break down the problem of *community interpretation* in two sub-problems: 1) finding an appropriate way of *representing* a community, and 2) taking advantage of this representation to identify the community most *characteristic features*. We solve the first sub-problem by representing a community as a set of sequences describing the evolution of its nodes. This encoding allows handling attributed dynamic networks, via their nodes topological measures and attributes. For the second sub-problem, we mine this set to identify sequential patterns fitting several criteria.

The process we propose includes 3 steps. The first is to identify a reference community structure, as explained in subsection A. In the second step, we search for emerging sequential patterns and extract the corresponding supporting nodes for each community. We explain the details of this process in subsection B. Finally, the third step is to choose the most representative patterns to characterize the communities according to various criteria that we explain in subsection C.

### A. Step 1: Detecting Communities

To detect how nodes evolve in terms of community membership, we need first a reference community structure. It would be possible to apply a dynamic method; however this results in complications due to the merging, splitting, disappearing and appearing of communities through time. For this reason, in this fist version of our tool, we decided to use only static communities. We detected them on the global weighted network $\mathcal{G}$ described in subsection II-A. Note that a similar approximation appears in [18]. To perform the detection, we apply the Louvain[9] algorithm, which is a two-phase hierarchical agglomerative approach. During the first phase, the algorithm applies a greedy optimization to identify the communities. During the second phase, it builds a new network whose nodes are the communities found during the first phase. Then, the process is repeated again iteratively.

### B. Step 2: Mining Emerging Sequences

We want to characterize each community according to the common evolution of the descriptors of its nodes over time. For this purpose, we need to identify series of descriptor values which appear often in the same community and over several time slices. This is precisely the goal of sequential pattern mining methods. Here, we describe the principle they rely upon, and how we take advantage of it.

We present an example network in Figure 1 to illustrate our method and the concepts it relies upon. It has 7 nodes whose interconnections change over 3 time slices. There is one attribute $a$ assigned to each node, whose value can also evolve through time. For the sake of simplicity, we only one topological measure: the degree.

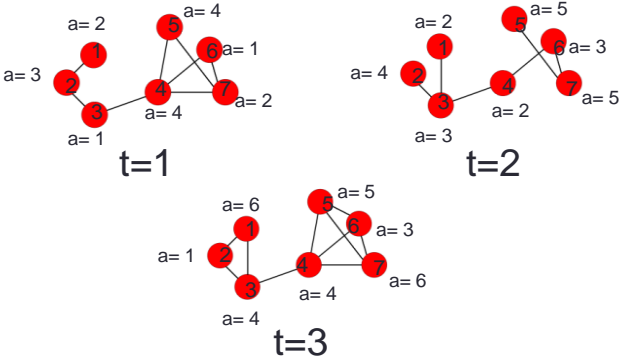

**Fig. 1.** Example Dynamic Network with 3 time slices, 7 nodes and 1 attribute

Let us first define the concepts necessary to the description of the method itself. An *item* $(D_i, x) \in D \times \mathfrak{D}_i$ is a couple constituted of a descriptor $D_i$ and a value $x$ from its domain $\mathfrak{D}_i$. The set of all items is noted $I$. The set of all possible items for the example network from Fig. 1 is $= \{a = 1, a = 2, a = 3, a = 4, a = 5, a = 6, d = 1, d = 2, d = 3, d = 4\}$, where $a$ is the only considered attribute and $d$ is the degree. An *itemset* $h$ is any subset of $I$. For example, $h = \{a = 1, d = 4\}$ is an itemset for our example network. A *sequence* $s = \langle h_1, \ldots, h_m \rangle$ is a chronologically ordered list of itemsets. The *size* $m$ of sequence $s$ is the number of itemsets it contains. For example, $\langle \{a = 1, d = 2\}\{a = 3, d = 3\} \rangle$ is a sequence of size 2 extracted from the network described in Fig. 1. A sequence $\alpha = \langle a_1, \ldots, a_\mu \rangle$ is a *sub-sequence* of another sequence $\beta = \langle b_1, \ldots, b_\nu \rangle$ iff $\exists i_1, i_2, \ldots, i_\mu$ such that $1 \leq i_1 < i_2 < \cdots < i_\mu \leq \nu$ and $a_1 \subseteq b_{i_1}, a_2 \subseteq b_{i_2}, \ldots, a_\mu \subseteq b_{i_\mu}$. This is noted $\alpha \sqsubseteq \beta$. It is also said that $\beta$ is a *super-sequence* of $\alpha$, which is noted $\beta \sqsupseteq \alpha$. An example of such relation for Fig. 1 is $\langle (a = 4)(d = 1) \rangle \sqsupseteq \langle (a = 4) \rangle$.

The *node sequence* of a node $v$ is a specific type of sequence noted $u(v) = \langle (l_{11}, \ldots, l_{k1}) \ldots (l_{1\theta}, \ldots, l_{k\theta}) \rangle$ where $l_{it}$ is the item containing the value of descriptor $D_i$ for $v$ at time $t$. A node sequence $u(v)$ includes $\theta$ itemsets, i.e. it represents all time slices. Each one of these itemsets contains all $k$ descriptor values for the considered node at the considered time. In other words, $u(v)$ contains all the available descriptor-related data for node $v$. These tuples will be used later to constitute the database analyzed by our method. As an example, the node sequence for node 1 in the network of Fig. 1 is $\langle (a = 2, d = 1)(a = 2, d = 1)(a = 6, d = 2) \rangle$.

The set of *supporting nodes* $S(s)$ of a sequence $s$ is defined as $S(s) = \{v \in V: u(v) \sqsupseteq s\}$. The *support* of a sequence $s$, $Sup(s) = |S(s)|/n$, is the proportion of nodes, in $\mathcal{G}$, whose node sequences are super-sequence of $s$. Similarly, The set of *supporting nodes* $S(s, C_c)$ of a sequence $s$ in $C_c$ is defined as $S(s, C_c) = \{v \in C_c: u(v) \sqsupseteq s\}$ and the *support* of a sequence in a community $C_c$, $Sup(s, C_c) = |S(s, C_c)|/|C_c|$, is the proportion of nodes in $C_c$, whose node sequences are super-sequence of $s$. Given a minimum support threshold noted $min_{sup}$, a *frequent sequential pattern (FS)* is a sequence whose support is greater or equal to $min_{sup}$. A *closed frequent sequential pattern* (CFS) is a FS which has no super-sequence possessing the same support.

In this study, we used the algorithm *CloSpan*[19] to find out all possible CFS for a given $min_{sup}$. It has two steps. At the first step, CloSpan creates the candidate set, which is a super-set of closed frequent sequences, and stores the elements into a so-called *prefix sequence lattice*. At the second step, the algorithm prunes this lattice, in order to eliminate non-closed sequences. This pruning technique relies on the fast subsumption checking method introduced by Zaki[20]. This technique manages a hash table in which the hash keys of a sequence is the sum of all the sequence ID's supportings that sequence.

CloSpan is an efficient algorithm, which can mine long sequences in practical time for real-world data. It outputs both the sequences and their supports, but not the supporting node sets, so an additional processing is required to identify them. In our case, we want to characterize communities in terms of CFS. Thus, we need to identify, for each community, its most representative sequential pattern(s). For this purpose, we turn to the notion of *emerging pattern*, i.e. a pattern more frequent in a part of the nodes than in the rest of it. The emergence of a pattern $s$ relatively to a community $C_c$ is measured by its *growth rate* given in Equation (7).

$$Gr(s, C_c) = \frac{Sup(s, C_c)}{Sup(s, \overline{C_c})} \qquad (7)$$

Here $\overline{C_c}$ is the complement of $C_c$ in $V$, i.e. $\overline{C_c} = V \setminus C_c$. The growth rate is the ratio of the support of $s$ in $C_c$ to the support of $s$ in $\overline{C_c}$. Therfore, a value larger than 1 means $s$ is particularly frequent (i.e. emerging) in $C_c$, when compared to the rest of the network. We consider that the higher the growth rate, and the more representative the sequence $s$ for community $C_c$.

In order to calculate the growth rate, it would be necessary to search CFS in all communities separately, which can be a costly operation. However, a more efficient method was proposed in[21] to handle the case where classes are assigned to item sequences. Our communities can be considered as classes, which is why the method is also relevant to our case. It is based on a modification of the analyzed data. First, let us note $\alpha \bullet C$ the concatenation of a sequence $\alpha = \langle a_1, \ldots, a_\mu \rangle$ and a symbol $C$, such that $\alpha \bullet C = \langle a_1, \ldots, a_\mu, \{C\} \rangle$. Instead of working on a sequence database constituted of $n$ tuples of the form $(v, u(v))$, we use a database $\mathcal{M} = (v, u(v) \bullet C(v))$, containing each node $v$, its node sequence $u(v)$ and community $C(v)$. Note that $u(v) \bullet C(v) = \langle h_1, \ldots, h_\theta, \{C(v)\} \rangle$, where $h_t$ is the itemset of $u(v)$ at time $t$. After having identified the frequent sequences by applying CloSpan to $\mathcal{M}$, the patterns concerning a community of interest $C_c$ can be obtained simply by selecting all the CFS ending with $C_c$. The support of such CFS (of the form $s \bullet C_c$) in $\mathcal{M}$ corresponds to the support one would have obtained on the non-concatenated database. Thus, all the necessary information to calculate $Gr$ is provided when applying CloSpan to $\mathcal{M}$.

Processing the growth rate of all CFS relatively to all communities then requires separating the CFS depending on how they end: we group together patterns related to the same community, and also those not related to any community. Let us note $r$ the number of CFS typically found for a community, as well as those found for the whole network. Indeed, the order of magnitude of these quantities is approximately the same. Then, we process each one of the $r$ patterns found in each one of the $\lambda$ communities. For such a pattern, we retrieve its support and that of the corresponding community-less pattern, as outputted by CloSpan.

Once the emerging CFS are identified for a community, we extract their supporting nodes, which are not directly outputted by CloSpan. To extract the supporting node set $S(s, C_c)$, for some specific pattern $s$ and community $C_c$, we use a naive approach consisting in accessing $\mathcal{M}$ and selecting the nodes whose sequences are the super sequences of $s$.

### C. Step 3: Selecting Sequential Patterns and Identifying Anomalies

After the emerging patterns are identified for a given community, together with their support, growth rate and supporting nodes, we need to select the most representative ones, in order to characterize the considered community. We give more attention to the most emerging pattern, i.e. the one whose growth rate is the highest. However, there is no guarantee for this pattern to cover a sufficient part of the community. And indeed, in practice it appears to be the opposite. It is thus needed to identify other complementary patterns, allowing us to obtain a more complete coverage of the community. Intuitively, we want to find a small number of patterns, such that they cover a significant part of the community, and are different in terms of supporting nodes. This amounts to defining the following constraints:

1. The intersection of the patterns supporting nodes sets must be minimal;
2. The union of these supporting nodes sets must be maximal (if possible: the whole community);
3. The number of patterns must be minimal.

Thus, the problem that we want to solve is to find the minimal number of patterns whose supporting nodes sets intersections is minimal while their union is maximal. Note that, in any case, we consider the patterns with the highest growth rate as the most emerging one, and search for additional ones to finish the coverage. In order to solve our problem, we select iteratively the most distant patterns, in terms of supporting node set. We use Jaccard's coefficient[22] as a distance measure between the node sets. In case of equality, the growth rate is considered as a secondary criterion. This iteration continues until it converges.

Besides considering patterns according to their growth rate, we also consider them according to their support, as a complementary analysis. For a given community, it is likely the highest supported patterns already include the majority of the nodes. So, unlike for growth rate-based patterns, it might not be necessary to apply the process designed to identify additional patterns. However, this cannot be guaranteed, since it depends on the considered network, attributes and topological measures. Supports are directly produced by CloSpan, so in order to select the highest supported patterns, we just need to analyze each pattern in each community.

Once the most characteristic patterns of a community have been identified (the most emerging one with its supplementary patterns, and the one with highest support), it is possible to use them to detect anomalies, i.e. nodes not following those patterns. Let $K_j(C_c)$ be the set of supporting nodes of pattern $j$ in community $C_c$, and $K(C_c) = \bigcup_{j \in \{1, \ldots, f\}} K_j(C_c)$ the supporting nodes for all representative patterns in the same community. Then we define the *anomaly nodes set* $E = C_c \setminus K(C_c)$ as the set of nodes not following any representative pattern. These nodes are different in the sense they do not follow the general trends of their communities. We detect anomalies automatically when finding representative patterns.

The overall complexity of our method includes calculating all topological measures, creating a global weighted network, applying the Louvain algorithm to detect the communities, applying the CloSpan algorithm to identify the patterns, processing their growth rates, and finally selecting the most representative ones. Among the considered topological measures, the transitivity has the highest complexity: $O(l^{3/2}\theta)$, for $l$ links and $\theta$ time slices. The processing of the global weighted network is in $O(l^2\theta)$. According to their respective authors, Louvain is in $O(n \log n)$[23] and CloSpan in $O(n^2)$[24], where $n$ is the number of nodes. However, the operations with the highest complexities are the processing of the growth rate and the selection of the most representative sequence, which are both in $O(\lambda r^2)$, where $\lambda$ and $r$ are the numbers of communities and of detected sequences, respectively. By definition, we have $\lambda \leq n$, and in practice, $r$ is generally much larger than $n$. Moreover, the number of time slices $\theta$ is smaller than both $n$ and $r$ by several orders of magnitude. Considering all simplifications and negligible terms, we get a total complexity of $O(\lambda r^2)$. In other words, the processing time mainly depends on the number of communities and sequences found in the data.

## IV. RESULTS

We now present the results obtained on real-world data. We selected the dynamic co-authorship network from[25], extracted from the DBLP database. Each one of the 2145 nodes represents an author.

Two nodes are connected if the corresponding authors published an article together. Each time slice corresponds to a period of five years. There are totally 10 time slices ranging from 1990 to 2012. The consecutive periods have a three year overlap for the sake of stability. For each author, at each time slice, the database provides the number of publications in 43 conferences and journals. We use this information to define 43 corresponding node attributes, and we add two more: the total number of conference and journal publications. Finally we have a total of 45 attributes. Our descriptors are these attributes and the topological measures described in subsection II-A.

The topological measures are discretized differently, depending on their nature. For node degree, we use the thresholds 3, 10 and 30. For transitivity, which is defined for $[0; 1]$, they are 0.35, 0.5, and 0.7. For embeddedness, which is also defined for $[0; 1]$, the intervals are 0.3 and 0.7. These intervals were determined to take into account distributions of these measures on the set of nodes and time slices: different thresholds correspond to areas of low density. For Guimerà et Amaral measures, we use the thresholds originally defined in[17], i.e., 2.5 for z et 0.05, 0.6, and 0.8 for P. The threshold used for z distinguishes community hubs ($z > 2.5$) and community non-hubs ($z \leq 2.5$). For the conference/journal publications, we consider the values $1, 2, 3, 4$ and $> 5$. For total journal or conference publications, we use the intervals of 5,10,20 and 50. These ranges are determined according to our knowledge of the domain.

After having applied Louvain, we found 127 communities in the global weighted network, for a modularity of 0.59. This value tells us that global weighted network is clearly modular. We discarded 96 of the communities, because they contain only one node. Amongst the remaining ones, 17 contain more than 10 nodes; the largest one having 335 nodes. We then searched the sequential patterns for these communities only, for a minimum support of 0.02. We could not execute the CloSpan algorithm for the smallest minimum supports, because of memory limitations. For each communities whose size is larger than 40, we find more than 5000 patterns. Most of these patterns include only topological measures.

### A. Most Supported Patterns

The most supported patterns are always a sequence of $z < 2.5$ for all communities, with changing sizes. This means the majority of the nodes for each community have the role of non-hub. As a reminder, Amaral&Guimerà define a community hub as a node whose internal degree is well above the average internal degree of its community[17]. Thus, the detected pattern means that the majority of the nodes are not particularly well-connected to their communities. Although this type of pattern appears in all communities, we can make a distinction in considering the size of the sequence.

In Table I, we list the size of the most supported sequential patterns with, for each one, its community label, community size and support. The communities whose sizes are between 39 and 45 (i.e. #40, 55 and 77) have long sequences (8, 7 and 7 resp.). Especially, the supports of communities 55 and 77 both reach the maximal value 1. This means in these communities, there is no remarkable hub author for a long time, or even if they appear sometime, they disappear very quickly. This observation is particularly interesting, and reflects the absence of a community leader who would structure the community through its many connections.

For community #115, the size of the sequence is 1, and its support is also 1. This means all the nodes which create this community had the role of non-hub together once, but for the rest of the time slices, they at least took the hub role once. For communities #38, 40 and 75, the support is less than 1, so we can say there is at least one hub, different from the rest of its community, and probably leading it. For communities #38, 40 and 75, the support is less than 1, meaning that an overwhelming majority of nodes plays the role of non-hub for long periods; however a small number of nodes take the place of hub possibly intermittently.

TABLE II
MOST SUPPORTED SEQUENCE SIZE FOR EACH COMMUNITY

| Commuity ID | Commuity Size | Sequence Size | Support Value |
|---|---|---|---|
| 38 | 335 | 2 | 0.99 |
| 40 | 43 | 8 | 0.97 |
| 42 | 109 | 5 | 1.00 |
| 45 | 227 | 3 | 1.00 |
| 55 | 39 | 7 | 1.00 |
| 61 | 204 | 3 | 1.00 |
| 75 | 140 | 4 | 0.99 |
| 77 | 41 | 7 | 1.00 |
| 86 | 111 | 3 | 1.00 |
| 98 | 113 | 5 | 1.00 |
| 106 | 134 | 5 | 1.00 |
| 115 | 125 | 1 | 1.00 |
| 125 | 79 | 3 | 1.00 |

We identified the authors who do not follow the most supported patterns for these 3 communities. For community #38, *Philip S. Yu*, *Jiawei Han* and *Beng Chin Ooi* are different from their communities. As expected, these nodes have a remarkably high number of connections within their communities, and the represented authors actually have leadership roles in their fields. Further analysis of the data also shows that they publish a total of more than 10 articles per time slice. In addition, they never took the non-hub role. Anomalies for communities #40 and 75 are respectively *Hans-*

*Peter Kriegel* and *Divesh Srivastava*. Here also, they are important authors in their community. Their sequences confirm that they are productive and do not take the non-hub role during all time slices.

### B. Most Emerging Patterns

For communities whose sizes are between 39 and 45, we do not find any emerging pattern containing a conference or journal. The most emerging patterns have a maximal growth rate of 1.79, which means there is no very distinctive sequential pattern for these communities. For the majority of the large communities, the most emerging pattern includes a specific conference or journal, which can be interpreted in terms of main theme of the community.

The other descriptors constituting the pattern are topological measures. As the most supported patterns, the item $z < 2.5$ appears the most often among the detected patterns. However, these most emerging patterns do not cover the majority of nodes. That is why, as we explained in subsection III-C, we looked for additional sequential patterns while minimizing the intersection of their supporting nodes with the previously chosen ones. These patterns generally consist of topological measures and do not have a very high growth rate. In the following part, we focus on the communities leading to the most interesting results. For each of them, we describe the most emerging pattern and present the anomalies. Each pattern is formally represented in brackets, as a sequence of itemsets which are represented between parentheses.

For community #61, the most emerging pattern is $<$ (ICML PUB. NUM=1) (DEGREE 3-10 $Z<2.5$)$>$, with growth rate 3.52 and support 0.30. This pattern refers to the authors who published once in ICML, then had a degree between 3 and 10 and became non-hubs. We extract 7 supplementary patterns to cover all the nodes of this community. Some of the interesting ones are $<$($Z<2.5$)( $Z<2.5$)( $Z<2.5$ CONF. PUB 1-5)(AAAI PUB 1)$>$ with growth rate 1.69 and support 0.30, and $<$(PART. COEFF 0.05-0.6 CIKM PUB. 1)$>$ with growth rate 1.40 and support 0.30. The former pattern refers to nodes that stay non-hub for a while, and then publish in conferences, before publishing in AAAI while losing their status of non-hub (without massively becoming hubs). The latter has no temporal dimension, but it shows the existence of nodes publishing in CIKM while having a peripheral position in the community, i.e. being significantly connected to other communities. The anomalies of this community are *Alex Alves Freitas*, *Claire Cardie*, *Edwin P. D. Pednault*. Among these authors *Alex Alves Freitas* does not have any publication for the first 8 time slices, before he starts publishing very efficiently in various conferences other than ICML or AAAI and journals. This can be interpreted as a Junior searcher progressively maturing. For the other two authors, while *Claire Cardie* publishes in ICML during the first 6 time slices at least once routinely, *Edwin P. D. Pednault* never published in not only ICML but also AAAI or CIKM.

The pattern $<$( PODS PUB 1)$>$ is the most emerging one in community #75. Its growth rate is 3.59 and its support is 0.40. This pattern shows that 40% of the authors of this community published at least once in PODS, which is a behavior significantly different from the rest of the network. There are 4 supplementary patterns to cover the rest of the community. These patterns refer to non-hub and peripheral nodes whose transitivity is very high, which means authors from this community tend to work in subgroups. The anomalies are *Ninghui Li*, *Feifei Li*, *Abdullah Mueen* who never published in PODS. The most emerging pattern of community 106 is $<$($Z<2.5$) ($Z<2.5$) ($Z<2.5$) ($Z<2.5$) ($Z<2.5$) ( PART. COEFF 0.05-0.6 KDD PUB. 1)$>$ with growth rate 2.87 and 0.40. This pattern refers to non-hub nodes staying non-hub for a while, then becoming peripheral nodes and publishing once in KDD. This evolution reflects a change in the community connectivity: nodes are at first loosely connected to other nodes in their own community, this overall internal connectivity improves, while the external connectivity (i.e. links with other communities) tend to become more heterogeneous. There are 4 supplementary patterns to cover the whole community. The supplementary patterns refer to the nodes with ultra-peripheral role, whose connections are usually inside their own community. Two anomalies of this community are *Stan Matwin* who is publishing in KDD more than one article routinely for every time slice, while not taking the non-hub role, and *Hua-Jun Zeng* who never publishes in KDD. In fact, *Hua-Jun Zeng*, while he does not produce any publication for the first 5 time slices, becomes very productive afterwards.

The most emerging pattern of community #45 is $<$(VLDB PUB. 3)( DEGREE 3-10 $Z<2.5$ )$>$ with growth rate 6.40 and support 0.30. This sequence tells us that there is a remarkable group of authors who published 3 times in the VLDB conference, before seing their degree reach a value between 3 and 10 and holding a non-hub role. There are 6 more sequential patterns that we have found to cover the rest of the community. One of them is $<$( $Z<2.5$ CONF. PUB 1-5)( $Z<2.5$ EMBED 0.3-0.7 ICDE PUB. 1 )$>$ with growth rate 2.30 and support 0.30. This pattern covers the non-hub nodes who published between 1 and 5 times in a conference, followed by being non-hub and having some connections outside of their community and publishing once in ICDE. The anomalies are *Ingmar Weber*, *Anastasia Ailamaki* who do not have any publication for the first 7 time slices, while they both become more and more productive for the last 3 time slices. Their publication number increases fast.

### C. Final Observations

To summarize our observations, the most emerging patterns in almost all communities usually include being non-hub and having a small number of publications in various journals or conference. Depending on the conferences or journals appearing in these patterns, it is possible to deduce the main theme of these communities. For some communities, however, the emerging sequential patterns are purely topological (no attributes). We can then assume that the members of these communities do not publish in a sufficiently homogeneous way so that it can appear under the form of patterns, which is itself a characteristic of the community. Another reason may simply be that the community members are connected to each other for different reasons than a common research theme (e.g. geographic or logistic constraints), in which case those do not appear in the attributes selected for our study. Regarding anomalies, one can distinguish different types of profiles. Some seem tocorrespond to authors whose main theme is different from that of the community in which they were placed. In some cases, we found out the authors had clearly moved to a different theme, or just started working in a given theme. They may also be authors active in another field, including conferences and journals not part of those used in the data we considered here. Another profile is that of junior researcher, whose number of publications and community position evolvjointly.These authors do not seem very active in their field in the first time slices. However, their number of publication and importance in their community increase with time.

## V. Conclusions

In this work, we tackled the problem of the characterization of communities in dynamic and attributed complex networks. We proposed a new representation of the information encoded in the network to store the topological information, the node attributes and the temporal dimension simultaneously. We used this representation to perform a search of emergingsequential patterns. Each community could then be characterized by its most distinctive patterns. We also took advantage of patterns to detect and characterize anomaly nodes in each community. We applied our method to a scientific collaboration network constructed from the public database DBLP. The results showed that our method is able to characterize the communities,in particular their research topic. The anomaly nodes we identified correspond to different types of profiles, such as community leaders, emerging researchers, or others changing research theme.

To our knowledge, this is the first formulation of the characterization of communities as a problem of data mining. Our goal was to overcome the limitations of the few existing studies[11, 13, 14] by proposing a systematic approach, taking into account the topologic structure, the nodal attributes and time. The representation of data we use has not been applied to the treatment of graphs before. The proposed process to extract the most relevant patterns based on a sequential pattern under constraint is original and we showed the consistency of interpretations with an application on a real-world network.

To limit the complexity of this first approach, we deliberately limited our analysis method by not considering the evolution of communities over time. In future works, we plan to take advantage of such communities, by inserting the appropriate information in the database used for the search patterns. We also plan to apply our method of analysis to other types of networks to explore its characterization capabilities. As another perspective, we can better use our representations of dynamic attributed network. Here we are only interested in mining emergingsequences. However, our data representation of the network can also be used to handle queriesconcerning the nodes,expressed in terms of topological measures or attributes. For instance, in our experiment, we saw that there were many nodes whose behavior was not typical of their community. Such queries could be used to studythemin further details,and better understand how they are different.